\documentclass[a4paper,11pt]{article} 
\usepackage{graphicx}
\usepackage{bbm}
\usepackage{amsmath}
\usepackage{mathrsfs}
\usepackage{color,soul}
\usepackage{geometry}
\usepackage{hyperref} 
\usepackage{amsmath,amssymb}

\usepackage{amsthm}

\usepackage{subfig}

\usepackage{multirow}
\usepackage{textcomp}
\usepackage{optidef}

\usepackage{algorithm}
\usepackage{algpseudocode}
\usepackage{authblk}

\begin{document}
\title{Energy storage in Madeira, Portugal: co-optimizing for arbitrage, self-sufficiency, peak shaving and energy backup}

\author[1]{Md Umar Hashmi}
\author[2]{Lucas Pereira}
\author[1]{Ana Bu\v{s}i\'c}
\affil[1]{{INRIA and DI ENS, ENS,}
		\textit{CNRS, PSL University}, Paris,
		France}
\affil[2]{ITI, LARSyS, Ténico Lisboa, and prsma.com}

\date{}
\maketitle

\begin{abstract}
Energy storage applications are explored from a prosumer (consumers with generation) perspective for the island of Madeira in Portugal. These applications could also be relevant to other power networks. 
We formulate a convex co-optimization problem for performing arbitrage under zero feed-in tariff, increasing self-sufficiency by increasing self-consumption of locally generated renewable energy, provide peak shaving and act as a backup power source during anticipated and scheduled power outages.
Using real data from Madeira we perform short and long time-scale simulations in order to select end-user contract which maximizes their gains considering storage degradation based on operational cycles.  
We observe energy storage ramping capability decides peak shaving potential, fast ramping batteries can significantly reduce peak demand charge.
The numerical experiment indicates that storage providing backup does not significantly reduce gains performing arbitrage and peak demand shaving. 
Furthermore, we also use AutoRegressive Moving Average (ARMA) forecasting along with Model Predictive Control (MPC) for real-time implementation of the proposed optimization problem in the presence of uncertainty.\\
\end{abstract}

\tableofcontents

\section{Introduction}
Medium sized isolated power networks often restrict the share of renewables and enforce stringent rules necessary to maintain safety and stability of the electrical power system. Authors in \cite{papadopoulos1991simulation} indicate that a high share of wind energy penetration could lead to large variations in active power generated due to sudden changes in wind speed. This in effect creates a huge mismatch between supply and demand, causing large variations in voltage and frequency leading to hazardous operating conditions. The enforcement regulations set by the Ministry of Industry and Energy of the Canary Government goes as far as restricting any further increase in wind farm installations directly connected to the power network. However, additional renewables are considered favourable if the generated energy is self-consumed locally \cite{calero2004action}.
{In \cite{bueno2006wind} the use of wind-powered hydro storage system for increasing the penetration of renewables is proposed for the island grid of Gran Canaria (part of Canarian Archipelago). In this solution, pumped storage acts as energy storage which is feasible due to the geography of the island and might not be applicable to similar isolated power networks. }

The work in \cite{vieira2017energy} presents a case study for Coimbra where residential solar and energy storage is 
locally consumed 
{with goal} of zero energy buildings. Authors observe that the electricity bill for the household reduces by more than 87\%. Furthermore, self-consuming intermittent renewable generation locally is desired by the utilities, as such generation makes load balancing, frequency and voltage regulation more challenging. Thus self-consuming renewable generation assists the power grid to accommodate a larger share of renewables \cite{luthander2015photovoltaic}.
Often utilities set the feed-in-tariff lower than the retail rate, making self-consumption more desirable for electricity consumers. 
Authors in \cite{bhandari2009grid} observe that by increasing self-consumption of renewables, their
financial feasibility
could be achieved a decade before in some European countries. 
Although
self-consumption holds multiple benefits for utilities, such a constraint creates a disparity for Distributed Generation (DG) owners, {as excess generation cannot be fed back to the grid}. 

In order to mitigate this disparity, these DG owners should co-optimize for additional revenue streams. In Germany DG owners until 2012 were incentivized for self-consumption rather than feeding power back to the grid \cite{moshovel2015analysis}. 
Under the case where DG owners are not allowed to supply power back to the grid, energy storage can facilitate load-shifting in real-time, minimizing consumption cost, increasing self-sufficiency \cite{luthander2015photovoltaic}. 
\textit{Self-sufficiency} is the ratio of total energy demand met by local generation and/or storage
with respect to cumulative energy needs.
Storage acts as a buffer of electrical energy and assists in the temporal shift of energy usage. {Authors in \cite{hill2012battery} describe the various applications for energy storage in future power networks.} Furthermore, with greater integration of intermittent renewables performing arbitrage and ancillary services will be more profitable \cite{hashmi2018effect}.
Co-optimizing energy storage for multiple applications has been proposed in many recent works. 
The authors in \cite{cheng2016co} co-optimize for arbitrage and frequency regulation, \cite{shi2018using} co-optimizes peak saving and frequency regulation. 
In this paper, we consider the case of Madeira Island, where utilities promote the inclusion of DG for self-consumption only. We propose the integration of a DG source along with energy storage battery. 
This battery facilitates self-consumption of locally generated energy, assist users in selecting a lower peak power contract, performing arbitrage and providing energy backup for instances of probable and scheduled power outages.\\

The key contributions of this paper are:\\
$\quad \bullet$ \textit{Co-optimization}: We propose a convex formulation for energy storage control for performing arbitrage, peak demand charge saving and backup reserve during power outages considering efficiency losses, ramping and capacity constraints for an energy storage battery.\\
$\quad \bullet$ {\textit{Storage profitability}: 
	The operational cycles govern storage degradation. The performance index for monetary value per cycle introduced in \cite{hashmi2018long} is used. This index indicates that battery is financially profitable in Madeira}.\\
$\quad \bullet$ \textit{Real time implementation}: We
use Auto-Regressive Moving Average (ARMA) processes to model temporal evolution in the MPC framework for real-time implementation considering uncertainty, motivated by our prior work \cite{hashmi2018netmetering}.\\

The key observations made from numerical results are:\\
$\quad \bullet$ 
Storage owners would benefit more by performing \textit{arbitrage}
with contracts with more price variations.\\
$\quad \bullet$ Ramping capability primarily decides the ability to reduce \textit{peak demand} charge savings for consumers.\\
$\quad \bullet$ 
{When DG generation is smaller than or approximately equal to inelastic load in magnitude then \textit{self-sufficiency} is governed by only the DG generator, otherwise, storage also contributes to it by increasing self-consumption.}\\
$\quad \bullet$ Providing \textit{energy backup} does not noticeably reduce storage ability to perform peak demand and/or arbitrage.
%

The paper is organized as follows. Section \ref{sectionii} introduces power system norms for consumers in Madeira. 
Section \ref{sectioniii} presents the system description.
Section \ref{sectioniv} formulates the co-optimization problem of performing arbitrage, peak shaving and energy backup.
Section \ref{sectionivplus} presets the real-time control of storage under uncertainty.
Section \ref{sectionv} presents the numerical results. 
Section \ref{sectionvi} concludes the paper.

\section{Power system norms in Madeira}
\label{sectionii}
Madeira is an archipelago in the North Atlantic Ocean, located about 1000 km southwest of mainland Portugal. It has a population of almost 270,000. 111,000 of which living in the capital city of Funchal. 
\subsection{Overview of the Madeira Electric Grid}
Madeira relies on local generation for electricity.
Empresa de Eletricidade da Madeira, S.A.
(\href{https://www.eem.pt}{EEM}) is the only DSO/TSO in Madeira, and is responsible for the activities related to production, transport, distribution and commercialization of electric energy. 
Energy generation sources used in Madeira are: thermal energy from fossil fuels like diesel and natural gas, hydro, wind, solid waste incineration (SWI), and photovoltaic (PV). 
In 2017, Madeira consumed 800GWh energy; thermal constituted about 70\% of the energy mix, with the remaining 30\% coming from renewable sources (hydro: 12.2\%, wind: 9.5\%, SWI: 4.8\%, and PV: 3.6\%) \cite{acif-ccim_madeira_2017}. The non-domestic sector (e.g., tourism and commerce), the domestic sector contributes 45\% and 30\% of total consumption respectively.
The remaining 25\% are contributed by public lighting (9\%), public buildings (8\%), industry (7\%), and agriculture ($<$ 1\%) \cite{acif-ccim_madeira_2017}. 
\subsection{Peak Power Contracts, Tariffs and Billing Cycles}
As of 2018, Low Voltage (LV) customers can select between 8 levels of peak power contract (PPC), three power tariffs, and two billing cycles \cite{vcharge_most_2018}.
Thus, there are total 48 different contracts that users can select from.
\begin{figure}[!htbp]
	\centering
	\includegraphics[width=5.9in]{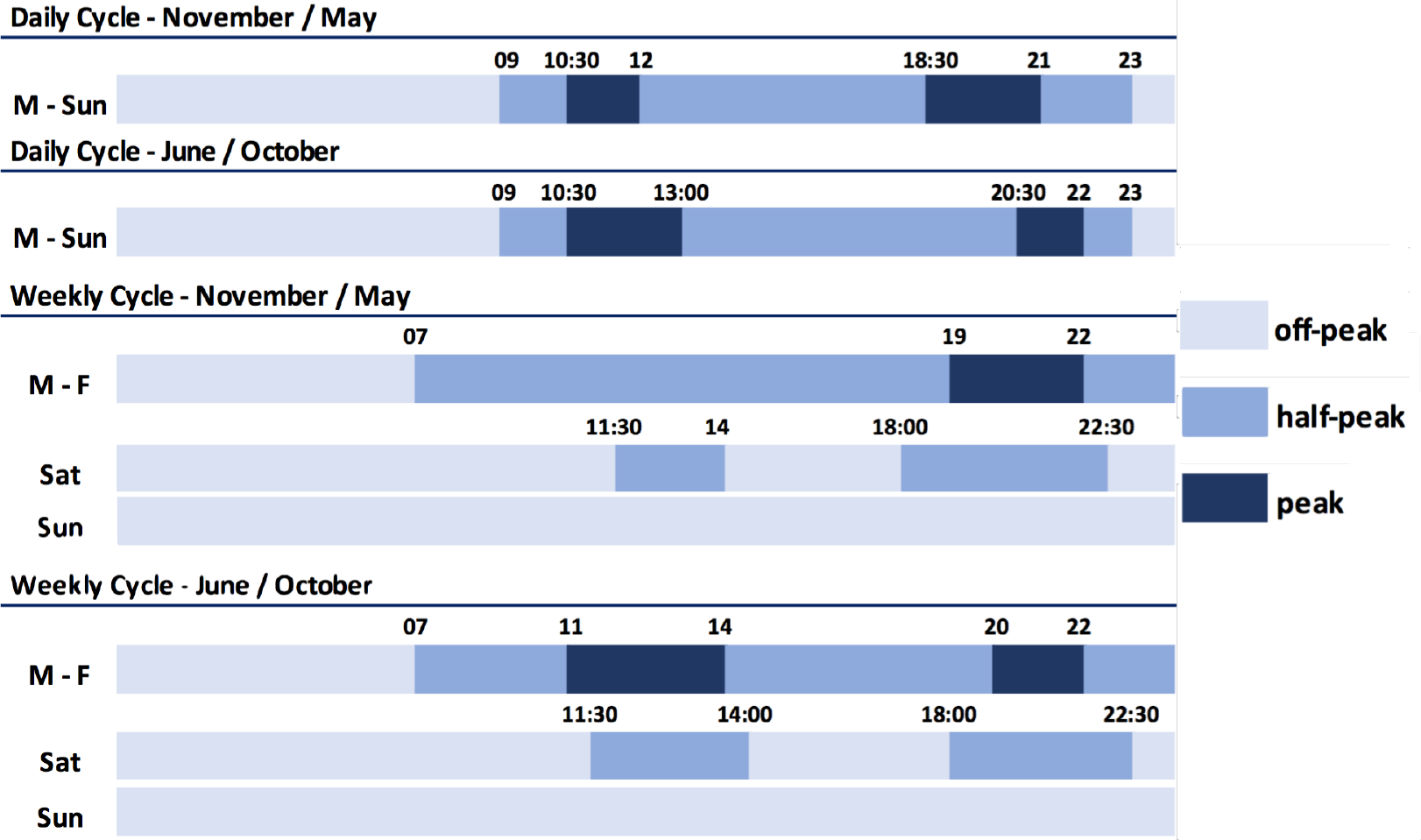}
	\caption{{Billing cycle schemes currently in place.}}
	\label{fig:daily_tariff}
\end{figure}
LV customers are  subject to a maximum peak power contract (kVA), ranging from 3.45 kVA to 20.70 kVA (see Tab.\ref{t:rates}). PPC value is selected by the customer based of their estimated electricity needs and should not be exceeded since the supply is shut-down when that happens. The disconnection of power is done locally and consumer can restart their energy-meter, however, sudden interruptions should be avoided as it may damage appliances. 
\begin{table}[!htbp]
	\caption {{Peak Power Contract for LV customer as of 2018.}}
	\label{t:rates}
	\begin{center}
		\begin{tabular}{| c| c| c|}
			\hline
			\multicolumn{3}{|c|}{Peak Power Contract (EUR/Day)}\\
			\hline
			PPC (kVA) & Single-rate (euros) & Dual/Triple-rate (euros) \\
			\hline
			3.45 & 0.1611  & 0.1643\\
			4.60 & 0.2096  & 0.2132\\
			5.75 & 0.2560  & 0.2590\\
			6.90 & 0.3040  & 0.3080\\
			10.35 & 0.4478 & 0.4532\\
			13.80 & 0.5902 & 0.5981\\
			17.25 & 0.7326 & 0.7436\\
			20.70 & 0.8751 & 0.8892\\
			\hline
		\end{tabular}
		\hfill\
	\end{center}
\end{table}
Regarding the energy tariffs, there are three options for LV customers.  \textit{Single}, \textit{dual} and \textit{triple-rate} {per kWh}. In the \textit{Single-rate} tariff the price is fixed at 0.1629 euros/kWh. In the \textit{dual-rate} tariff there are two \textit{time-of-use} (ToU) prices (\textit{peak} at 0.1894 euros, and \textit{off-peak} at 0.0982 euros), whereas in the \textit{triple-rate} tariffs have three ToU price levels (\textit{peak} at 0.2153 euros, \textit{off-peak} at 0.0982 euros, and \textit{half-peak} at 0.1716 euros).
The \textit{off-peak}, \textit{peak} and \textit{half-peak} periods are defined in advance based on the notion of \textit{daily} and \textit{weekly} billing cycles. In the \textit{daily} cycle there are no distinctions between weekend and workdays, whereas in the \textit{weekly cycle} there are different \textit{time-of-use} periods for work-days, Saturdays and Sundays. Fig.~\ref{fig:daily_tariff} summarizes the billing cycle schemes in practice as of 2018. For \textit{dual-rate tariff}, the \textit{peak} period includes also the \textit{half-peak} period.

Consumer electricity bill consists of two components: a fixed component that depends on the contracted power (kVA), and a dynamic component governed by actual energy consumption (kWh).
{Note that the fixed component is governed by apparent power: a function of active and reactive power. In this work, we consider PPC levels to a function of only active power as apparent power is primarily governed by active power. Reactive power compensations as proposed in \cite{hashmi2019arbitrage} could be applied for the storage control in this work.} 
\subsection{Self-Consumption and Renewables in Madeira}
In Madeira island, since 2014, new mini and micro-producers are not allowed to feed-in excess production to the local grid, thus excess generation is wasted \cite{acif-ccim_madeira_2017}.
Due to such a constraint DGs are sized to maximize self-consumption and minimize excess production \cite{prsma_data_2018}.
Counterintuitively, in a period that one should expect an explosion in the number of micro-producers leveraged by the relatively low prices of solar PV technologies, Madeira is experiencing an {stagnation} on the number of new solar PV installations due to the norms set by EEM.
The main reason for this change in the local legislation is to protect the grid from the issues associated with the intermittent and uncertain nature of renewable production from solar in a total energy system. 
In case of the electrical LV networks in the rural areas of Madeira Island which are at the edge of the radial distribution network, when associated with low consumption and high production periods, it is very likely to observe the phenomena of voltage increase
\cite{acif-ccim_madeira_2017}. 
%
%
%
%
\section{System Description}
\label{sectioniii}
In this work we consider a prosumer with inelastic load and rooftop solar generation as DG and 
an energy storage battery. 
The battery serves four purposes:
(i) increase self-consumption, thus reducing waste of excess generation if any,
(ii) perform energy arbitrage,
(iii) minimize the peak demand charge and
(iv) maintain battery charge level for scheduled and/or anticipated power failures.\\

\textbf{Notation:}
We consider operation over a total duration $T$, divided into $N$ steps indexed by $\{1,...,N\}$. The duration of each step is denoted as $h$. Hence, $T=hN$. 
At time instant $i$, the information available is the end user consumption $d_i$, the renewable generation ${r}_i$ and the storage energy output $s_i$. 
The load without storage is denoted as
$
z_i = d_i - r_i.
$
The load seen by grid is denoted as
$
L_i = d_i - r_i + s_i.
$
\\

\textbf{Battery Model}
The efficiency of charging and discharging 
is denoted by $\eta_{ch}, \eta_{dis} \in (0,1]$, respectively.
The change in energy level of the battery is denoted as 
$x_i$=$h \delta_i$, where $\delta_i$ denotes storage ramp rate at time instant $i$,
%
such that $\delta_i \in [\delta_{\min}, \delta_{\max}], \forall i$ and $\delta_{\min} \leq 0$ and $\delta_{\max} \geq 0$ are the minimum and maximum ramp rates (kW); $\delta_i >0$ implies charging and $\delta_i <0$ implies discharging. {The energy output of storage in the $i^{\text{th}}$ instant is given by $s_i = \frac{[x_i]^+}{\eta_{ch}} - \eta_{dis} [x_i]^-$, where $[x]^+$=$\max(0, x)$ and $[x]^-$=$-\min(0, x)$.} 
The ramping constraint induce limits on $s_i$ given by
\begin{equation}
s_i \in [\delta_{\min}h\eta_{dis}, \delta_{\max}h/\eta_{ch}], \quad \forall i.
\label{constraintramp}
\end{equation}
The energy stored in the battery is denoted as $b_i$, defined as $b_i = b_{i-1}+ x_i$. Battery capacity constraint is given as
\begin{equation}
b_i \in [b_{\min}, b_{\max}], \quad \forall i. 
\label{constraintcapacity}
\end{equation}
where $b_{\min}$ and $b_{\max}$ are minimum and maximum permissible battery charge levels respectively. We use xC-yC notation to represent the relationship between ramp rate and battery capacity. xC-yC implies battery takes 1/x hours to charge and 1/y hours to discharge completely.
%
\section{Co-optimizing energy storage}
\label{sectioniv}
Optimizing the energy storage is essential due to its high cost. 
In context of our present work we do not evaluate the fixed electricity cost as under such a case storage can only be used either for backing up excess generation and/or for peak demand shaving.
Backing up energy will require no look-ahead and greedy behavior leads to optimality \cite{su2013modeling}. The optimal solution in such a case is governed by the sign of $z_i$ and is given as
\begin{itemize}
	\item if ${z}_i \geq 0$ then battery should discharge such that \\
	$s_i = \max\left\{- {z}_i, \delta_{\min}h\eta_{dis}, ({b_{i-1} - b_{\min}}){\eta_{dis}}  \right\} $
	\item if ${z}_i < 0$ then battery should charge such that \\
	$s_i = \min\left\{-{z}_i, \delta_{\max}h /\eta_{ch},
	({b_{\max} - b_{i-1}})/{\eta_{ch}}
	\right\} $.
\end{itemize}
Next we formulate co-optimization problem for storage control for arbitrage, peak shaving and energy backup.
\subsection{ToU pricing + zero feed-in-tariff + Peak-Shaving}
The optimal arbitrage problem with battery ($P_{arb}$) is defined as the minimization of the cost of total energy consumed denoted as $ \min \sum_{i=1}^N [L_i]^+ p_{\text{elec}}(i)$ subject to the battery constraints. $p_{\text{elec}}(i)$ denotes the electricity price for consuming electricity, i.e. $L_i >0$, for instant $i$. Here we assume the feed-in-tariff to be zero. This optimization framework is a special case of the problem studied in \cite{hashmi2017optimal}, \cite{cruise2014optimal}. 
Under zero feed-in-tariff, only consumed energy is charged and end-user gets no incentive in supplying power back to the grid (i.e. $L_i < 0$). 
This is emulated using a variable $\theta_i = \max(0, L_i)$.
Keeping $L_i \geq 0$ will maximize the self-consumption of renewable generation locally.

In this formulation we also consider peak demand shaving by reducing the peak demand contract the end-user should opt for. Consider the maximum demand of the user without storage is $P_{\max}$ then the inclusion of energy storage would bring this maximum demand lower proportional to its ramp rate, considering look-ahead optimization ensures battery has enough capacity to discharge during peak demand.
The end user operates the energy storage for minimizing the cost of consumption, increasing self-consumption by reducing the waste of excess of generation and restraining peak demand. This optimization problem 
is given as

\begin{mini!}[1]
	{s_i}{ \sum_{i=1}^N {p}_{\text{elec}}(i)\theta_i h}{}{(P_{opt})~~}
	\label{eq:optprob1}
	\addConstraint{\text{Ramping constraint,~} }{Eq.~\ref{constraintramp}}
	\addConstraint{\text{Capacity constraint,~}}{Eq.~\ref{constraintcapacity}}
	\addConstraint{\text{Self-sufficiency,}~~\theta_i }{\geq 0}
	\addConstraint{\text{Arbitrage,}~~\theta_i }{\geq [z_i + s_i]}
	\addConstraint{\text{Peak shaving,}~~[z_i + s_i]/h }{\leq P_{\max}^{set}}
\end{mini!}

The peak power threshold, $P_{\max}^{set}$, is selected close to the power level ($P_{\max} + \delta_{\min}$), subject to $P_{\max}^{set} \geq (P_{\max} + \delta_{\min})$. $P_{\max}^{set}$ is selected by the electricity consumer as a PPC contract with the utility in Madeira. 

\subsection{Storage for BackUp with Arbitrage + Peak Shaving}
The valuation of energy storage devices performing energy backup is hard to quantify and often ignored in assessing the value of energy storage devices. Regions where power network is not very reliable, consumers install energy storage devices and a power converter to charge the battery while power is available through the grid and immediately start discharging when the grid supply is not active. Installing such devices provides uninterrupted power supply to the user, enhancing the reliability of power supply at the consumer end. Abrupt disconnection of power supply drastically affects the life of certain appliances. 
Energy storage is also used for energy backup in developed countries; loads like data centers and hospitals are critical and therefore, local backup of energy is essential. Here we consider two types of backup modes:\\
$\quad \bullet$  \textit{Pre-scheduled unavailability of power}: Due to scheduled maintenance power outages could occur. 
It is essential to consider such incidents in case of Madeira as being an isolated power network it has less inertia and less redundancy making it more prone to scheduled outages.
In such cases the time of the outage denoted as $i_{\text{incident}}$, is known a priori and users can maintain the battery level above $b_{\text{set}}$, so as in absence of grid supply users can meet its energy needs. This is represented as an additional constraint denoted as $b_i= b_{i_{\text{incident}}}\geq  b_{\text{set}}$.\\
$\quad \bullet$  \textit{Probability of power loss}: based on past failure incidents a probability of power failure can be calculated; for instance, the chances of power failure due to load shedding are much more probable during morning and evening peak than any other time of the day. $\mathbb{P}_i$ denotes the probability of power failure during the instant $i$. If $\mathbb{P}_i$ is high than the user should maintain a greater charge level in the battery. The co-optimization problem combined with planned and probable outages is given as follows

\begin{mini!}[1]
	{s_i}{\sum_{i=1}^N {p}_{\text{elec}}(i)\theta_i h - \lambda \mathbb{P}_i b_i }{}{(P_{\text{madeira}})~~}
	\label{eq:optprob2}
	\addConstraint{\text{Ramping constraint,~}}{Eq.~\ref{constraintramp}}
	\addConstraint{\text{Capacity constraint,~}}{Eq.~\ref{constraintcapacity}}
	\addConstraint{\text{Self-sufficiency,}~~\theta_i }{\geq 0}
	\addConstraint{\text{Arbitrage,}~~\theta_i }{\geq [z_i + s_i]}
	\addConstraint{\text{Peak shaving,}~~[z_i + s_i]/h }{\leq P_{\max}^{set}}
	\addConstraint{\text{Backup for probable outage,}~~b_i }{= b_{i_{\text{incident}}}\geq  b_{\text{set}}}
\end{mini!}

where $\lambda$ is a scaling factor.
%
Note $P_{opt}$ and $P_{\text{madeira}}$ are convex in nature as the objective function and associated constraints are convex.	

\subsection{Open Source Codes}
The co-optimization formulations and benchmarks presented in this chapter is made open source. The link for the code is ~ \url{https://github.com/umar-hashmi/MadeiraStorage}.

\section{Real-time Control under Uncertainty}
\label{sectionivplus}
The decision variables for the optimization problem $P_{\text{opt}}$ and $P_{\text{madeira}}$ is the energy storage output $s_i$. The stochastic variable in these settings is the the net-load excluding energy storage output, $z_i$. We use AutoRegressive Moving Average (ARMA) forecasting for modeling $z_i$. The details of the model is described in Section~\ref{armamodel} and model predictive control algorithm is described in Section~\ref{secmpc}.
\subsection{Modeling Uncertainty: ARMA Forecasting}
\label{armamodel}
We define the mean behavior of past values of net load without storage at time step $i$ as
\begin{equation}
{\bar{z}}_i=\frac{1}{D} \sum_{p=1}^D z_{i-pN} \quad \forall i \in \{k,...,N\}, k\geq 1,
\end{equation}
where $D$ is the number of days in the past whose values are considered in calculating $\bar{z}$.
The forecasted net load given as
$
\hat{z}_i= \bar{z}_i + \hat{X}_i   \quad \forall i \in \{k,...,N\}, k\geq 1,
\label{zhat}
$
where $\hat{X}_i$ represents the forecasted difference from the mean behavior. We define $\hat{X}_i \quad \forall i \in \{k,...,N\}$ as
\begin{equation}
\hat{X}_\text{k}=\alpha_1 \hat{X}_{\text{k}-1} + \alpha_2 \hat{X}_{\text{k}-2} + \alpha_3 \hat{X}_{\text{k}-3} + \beta_1 \delta_\text{k}^1+\beta_2 \delta_\text{k}^2+\beta_3 \delta_\text{k}^3,
\label{xhatk}
\end{equation}
where $\delta_k^m=(z_{k-mN} - {\bar{z}}_{k-mN})$ and $\alpha_i, \beta_i \forall i \in\{1,2,3\}$ are constant. We use the errors in net load without storage for the past three time steps and the error in the same time step for past three days. At time step $i=k-1$ we calculate $\hat{X}_k$ as shown in Eq~\ref{xhatk}. 
Using $\hat{X}$ we calculate $\hat{z}$.
\subsection{Model Predictive Control}
\label{secmpc}
The vector $\hat{z}$ for instants $i$ to $N$ is fed to MPC for calculating optimal energy storage actions for time step $i$. Similar steps are done for $i \in\{k+1,...,N\}$, till the end of time horizon is reached. Real-time algorithm is presented as \texttt{ForecastPlusMPC}.
\begin{algorithm}
	{\textbf{Inputs}: {$\eta_{\text{ch}}, \eta_{\text{dis}}, \delta_{\max}, \delta_{\min}, b_{\max}, b_{\min}$}, $b_0$, ${p}_{elec}$, $h, N, T,i=0$}\\
	\begin{algorithmic}[1]
		\While{$i < N$}
		\State Increment $i$=$i+1$ and forecast $\hat{z}$ from time step $i$ to $ N$
		\State Solve for $s^* = P_{\text{madeira}}({p}_{elec}, \hat{z}, h, N, T)$
		\State $b_i^*= b_{i-1}+[s^*]^+\eta_{ch} - [s^*]^-/\eta_{dis}$ and Update $b_0=b_i^*$
		\EndWhile
	\end{algorithmic}
	\caption{\texttt{ForecastPlusMPC}}\label{alg:3}
\end{algorithm}

\begin{figure}[!htbp]
	\center
	\includegraphics[width=6.5in]{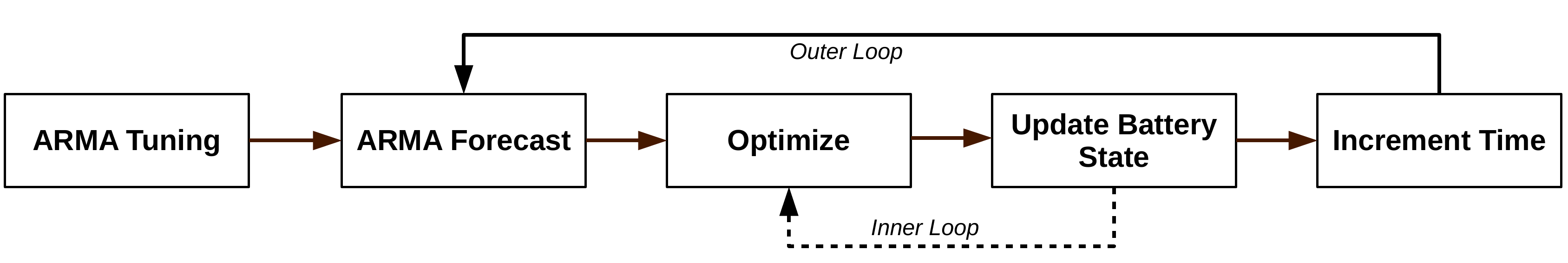}
	\caption{{Receding horizon MPC with forecasting}}
	\label{mpcdf}
\end{figure}
\section{Numerical Results}
\label{sectionv}
For the numerical evaluation we use a battery with initial charge level, $b_0$=1kWh, $b_{\max}$=2kWh, $b_{\min}$=0.2kWh, $\eta_{ch}$=$\eta_{dis}$=0.95. The home has 6.25 kWp solar PV. The performance indices used for evaluating simulations are: 
$\quad \bullet$  \textit{Arbitrage Gains} ($G_{arb}$), 
$\quad \bullet$  \textit{Peak shaving gains} ($G_{peak}$): difference between nominal Peak Power Contract (PPC) and the new PPC contract after adding storage.\\
$\quad \bullet$  \textit{Self-sufficiency} (SS): calculated using
total energy consumed, PV generation and storage output.\\
$\quad \bullet$  \textit{Gains per cycle}: In our prior work \cite{hashmi2018long} we develop a mechanism to measure the number of cycles of operation based on depth-of-discharge of energy storage operational cycles. We use total gains, $G_T$=$G_{arb}$+$G_{peak}$, to calculate euros/cycles gained by operating energy storage as one of the performance index.
This index puts a financial value to operational cycles of the battery.

We perform deterministic simulations for arbitrage and peak demand shaving in Section~\ref{sixA}. Section~\ref{sixB} presents numerical results for energy storage performing backup along with arbitrage and peak shaving. Section~\ref{sixC} compares forecast plus MPC with results for a week with respect to the deterministic results.

\subsection{Deterministic Solution for $P_{opt}$}
\label{sixA}
We compare various pricing contracts and propose the best contract that energy storage owners can select for maximizing their gains. Here we consider only the daily cycles since this is the most commonly selected option in Madeira. Two types of simulations results are presented: simulation on a shorter-time scale, (i.e., for a day) and for longer time-scale, (i.e., for a month). {The load data is collected from a facility in Madeira. } Single-rate tariff is used as the nominal case with respect to which profit and performance improvements are calculated.
\subsubsection{Shorter Time Scale: A day}
Simulations using 2 and 3-level ToU price for 4 different batteries are conducted for load data of 18th May, 2018. 
\begin{figure}[!htbp]
	\center
	\includegraphics[width=5.5in]{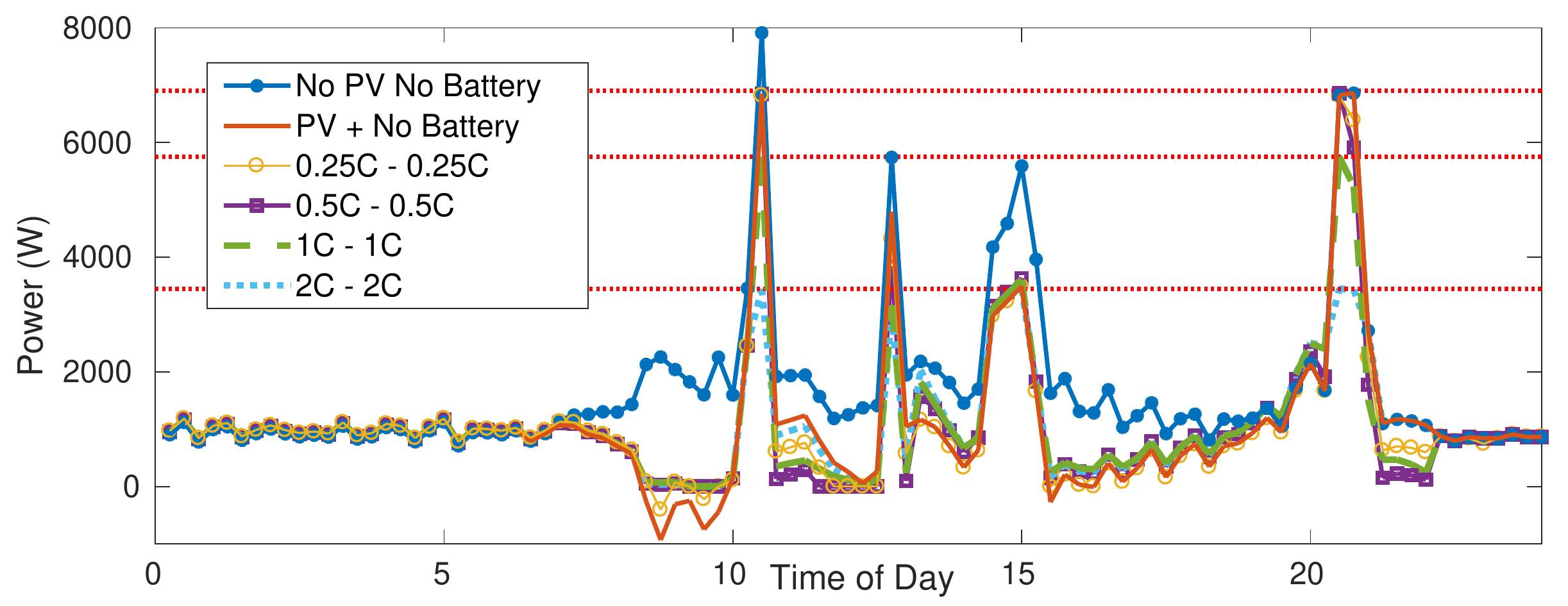}
	\caption{{Net Load with/without solar and with/without battery for $h=0.25$ hours}; here we consider \textit{4 different batteries}}
	\label{netload}
\end{figure}
From Table~\ref{resulttab1} we can conclude that 3-level ToU provides higher gains for end-user, fast ramping battery can increase $G_{peak}$, however, $G_{arb}$ deteriorates due to greater contribution of the battery performing peak reduction. Fig.~\ref{netload} shows the variation of net load for 3-level ToU. 
\begin{table}[!tbph]
	\caption {Comparison for 1 day}
	\label{resulttab1}
	\begin{center}
		\begin{tabular}{| c| c| c|c| c|c|c|}
			\hline
			\multirow{2}{*}{Case} & $G_{arb}$ & PPC & $G_{peak}$  & SS & $G_{T}$ & \textbf{euros/cyc} \\ 
			& euros & kVA  & euros	     & \%       & euros & \\
			\hline
			\multicolumn{7}{|c|}{No Battery} \\
			\hline
			No PV & - &10.35 & - & - & -&- \\
			PV  & - &6.9 & 0.144 &32.9& 0.144 & - \\
			\hline
			\multicolumn{7}{|c|}{2 Level ToU with Battery} \\
			\hline
			0.25C-0.25C 	& \textbf{0.240} & 6.9 & 0.144 & 36.2&{0.384} & 0.568  \\
			0.5C-0.5C	& 0.235 & 6.9 & 0.144 & 36.6&{0.379} & 0.561  \\
			1C-1C		& 0.235 & 5.75 & 0.192 & 36.6 &{0.427} & \textbf{0.631} \\
			2C-2C		& 0.212 &\textbf{ 3.45} & \textbf{0.287} & 36.3&\textbf{0.499} & 0.363  \\
			\hline
			\multicolumn{7}{|c|}{3 Level ToU with Battery}\\
			\hline
			0.25C-0.25C 	& 0.333 & 6.9 & 0.144 & 36.2&{0.477} & \textbf{0.704}  \\
			0.5C-0.5C	& \textbf{0.366} & 6.9 & 0.144 & 36.2&{0.510} & 0.376  \\
			1C-1C		& 0.351 & 5.75 & 0.192 & 36.1&{0.543} & 0.346  \\
			2C-2C		& 0.301 & \textbf{3.45} & \textbf{0.287} & 36.1&\textbf{0.587} & 0.379  \\
			\hline
		\end{tabular}
		\hfill\
	\end{center}
\end{table}
\begin{table}[!tbph]
	\caption {Comparison for Longer Time Scale}
	\label{resulttab2}
	\begin{center}
		\begin{tabular}{| c| c| c|c| c|c|c|}
			\hline
			\multirow{2}{*}{Case} & $G_{arb}$ & PPC & $G_{peak}$  & SS & $G_{T}$ & \textbf{euros/cyc} \\ 
			& euros & kVA  & euros	     & \%       & euros	 &  \\
			\hline
			\multicolumn{7}{|c|}{No Battery} \\
			\hline
			No PV& - &17.25 & - & - & -&- \\
			PV & - &17.25 & 0 & 34.1 & 0&- \\
			\hline
			\multicolumn{7}{|c|}{2 Level ToU with Battery} \\
			\hline
			0.25C-0.25C 	& 18.38 & 13.8 & 4.27 & 34.9 &22.65 & \textbf{0.855} \\
			0.5C-0.5C	& 18.38 & 13.8 & 4.27 & 34.9 &22.65 & 0.855  \\
			1C-1C		& 18.38 & 13.8 & 4.27 & 34.9 &22.65 & 0.855 \\
			2C-2C		& 18.37 & \textbf{10.35} & \textbf{8.54} & 34.9 &\textbf{26.92} & 0.669 \\
			\hline
			\multicolumn{7}{|c|}{3 Level ToU with Battery} \\
			\hline
			0.25C-0.25C 	& 25.01 & 13.8 & 4.27  & 34.9&29.28 & \textbf{1.035} \\
			0.5C-0.5C	& 25.89 & 13.8 & 4.27  & 34.7 &30.16 & 0.640 \\
			1C-1C		& \textbf{26.08} & 13.8 & 4.27 & 34.7 &30.36 & 0.589 \\
			2C-2C		& 26.06 &\textbf{10.35} & \textbf{8.54} & 34.6 &\textbf{34.60} & 0.568  \\
			\hline
		\end{tabular}
		\hfill\
	\end{center}
\end{table}
Integration of storage leads to approximately 3\% saving of electricity bills. Prior to installation of storage, 5.8\% of solar generation was wasted and with addition of storage the waste is reduced to zero. 
The simulations also shows that additional gains are possible by reducing the PPC shown in Fig~\ref{netload} and Table~\ref{resulttab1}.
\subsubsection{Longer Time-Scale: A month}
\begin{figure}[!htbp]
	\center
	\includegraphics[width=6in]{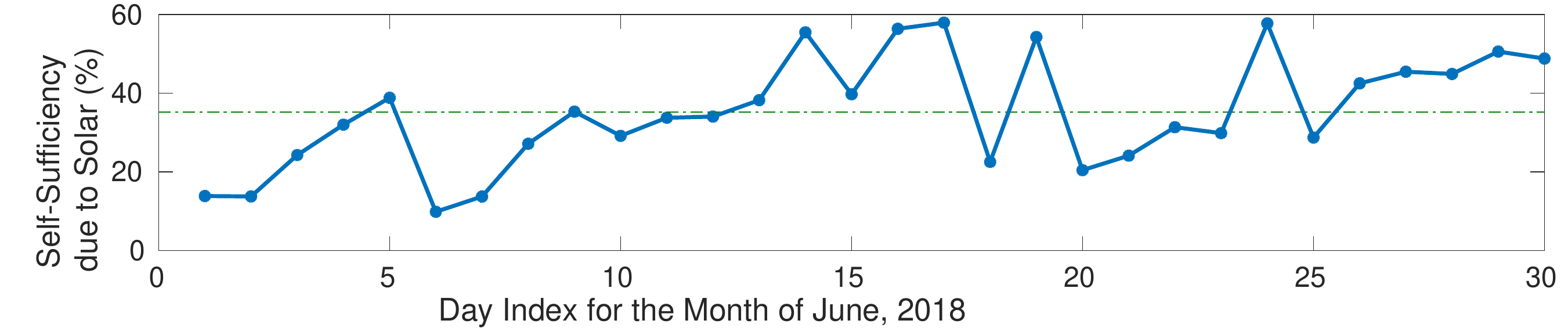}
	\caption{{Self-Sufficiency due to Solar for June 2018}}
	\label{ss}
\end{figure}

Longer time scale simulations are conducted for the month of June, 2018. Table~\ref{resulttab2} shows that energy storage does not contribute significantly towards self-sufficiency. Fig.~\ref{ss} shows the variation of SS for each day due to PV.
Long time scale simulations also indicate that 3 level ToU is more beneficial. 
{Further the gains per cycle indicate that energy storage is highly profitable. A typical LiIon battery costing around euros 500/kWh could perform around 4000 cycles at 100\% DoD. Thus such a battery (of 2kWh size) to be profitable should make more than euros 0.25/cycle.  Table~\ref{resulttab1} and Table~\ref{resulttab2} show that euros/cycle is significantly more than euros 0.25.}	
\subsection{Co-optimizing with Power Backup}
\label{sixB}
\begin{figure}[!htbp]
	\center
	\includegraphics[width=5.5in]{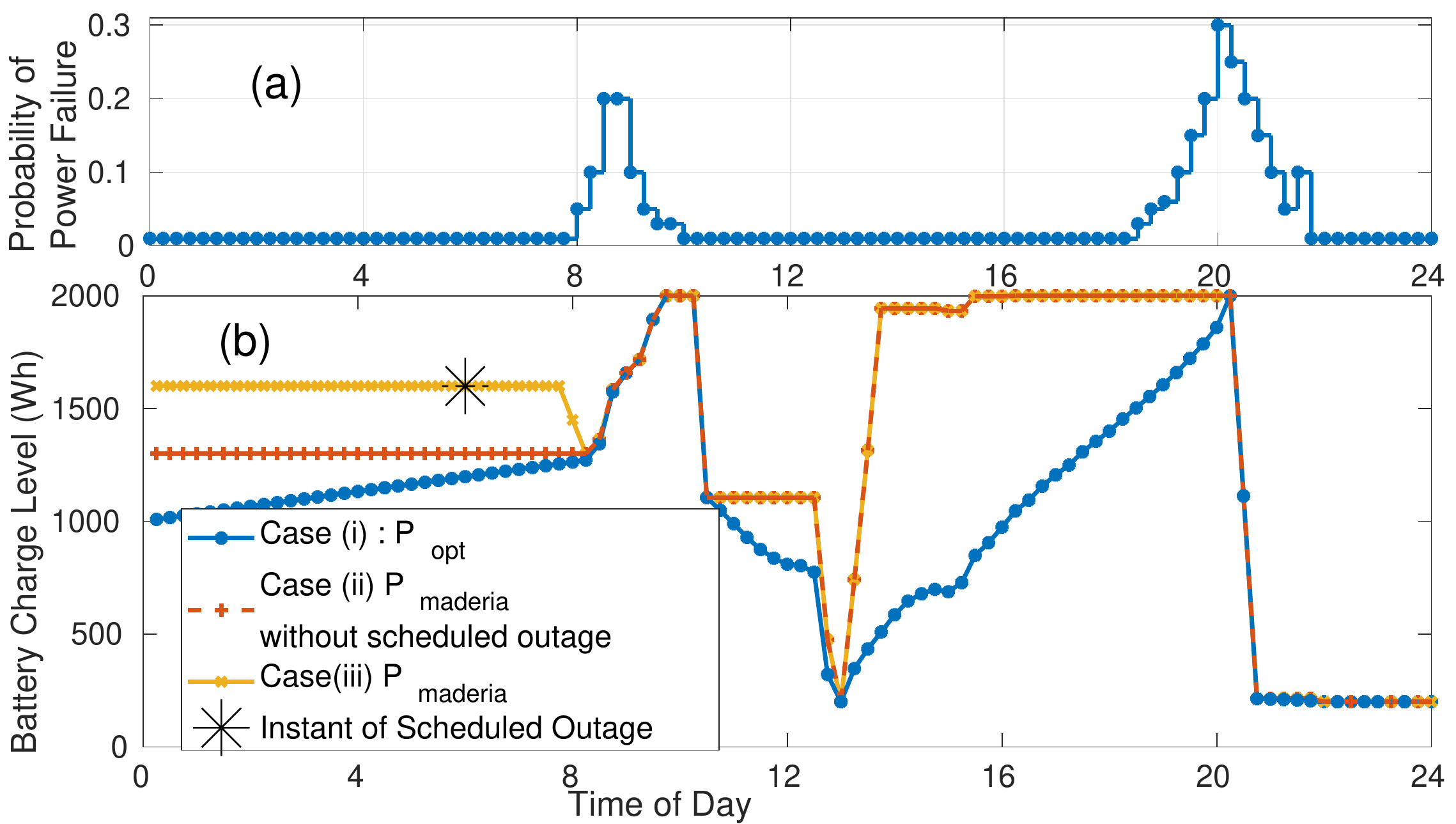}
	\caption{{(a) Probability of power failure, (b) Battery charge level}}
	\label{prob}
\end{figure}

The probability of power failure used for scheduling energy storage backup is shown in Fig.~\ref{prob}(a). The probability of power failure on a typical day is primarily because of load-shedding, which happens more during peak consumption hours.
We assume that there is a scheduled power outage incident at $i_{\text{incident}}$=6 am. During this time storage should maintain an state of charge of 80\% or higher, therefore, $b_{set} = 0.8 b_{\max}$.
\begin{table}[!htbp]
	\caption {{Comparison of gains for power backup}}
	\label{backup_tab1}
	\begin{center}
		\begin{tabular}{| c | c| c| c|}
			\hline
			Index& Case (i) $P_{opt}$& Case (ii) &Case (iii) $P_{madeira}$ \\ 
			\hline
			$G_{arb}$ & 0.3006 & 0.3006 & 0.2976\\
			$G_{peak}$ & 0.2867 & 0.2867 & 0.2867\\
			\hline
		\end{tabular}
		\hfill\
	\end{center}
\end{table}
Fig.~\ref{prob}(b) shows battery charge level for three cases: (i) $P_{opt}$: arbitrage with peak shaving, (ii) Arbitrage with peak shaving and backup for probable outage governed by failure probability in Fig.~\ref{prob}(a) and (iii) $P_{madeira}$: arbitrage, peak shaving, probable outage with scheduled outage. As evident from Fig.~\ref{prob}(b) the charge level for Case (iii) maintains a higher charge level during $i_{\text{incident}}$. For case (ii) and case (iii) maintains a high charge level during probable outage during morning and evening peak.
As shown in Table~\ref{backup_tab1}, the effect on gains due to performing backup is insignificant, less than 1\% in this case. 
For these cases self-sufficiency remains fairly similar.
\subsection{Real-Time Implementation (Forecast plus MPC)}
\label{sixC}
\begin{figure}[!htbp]
	\center
	\includegraphics[width=6in]{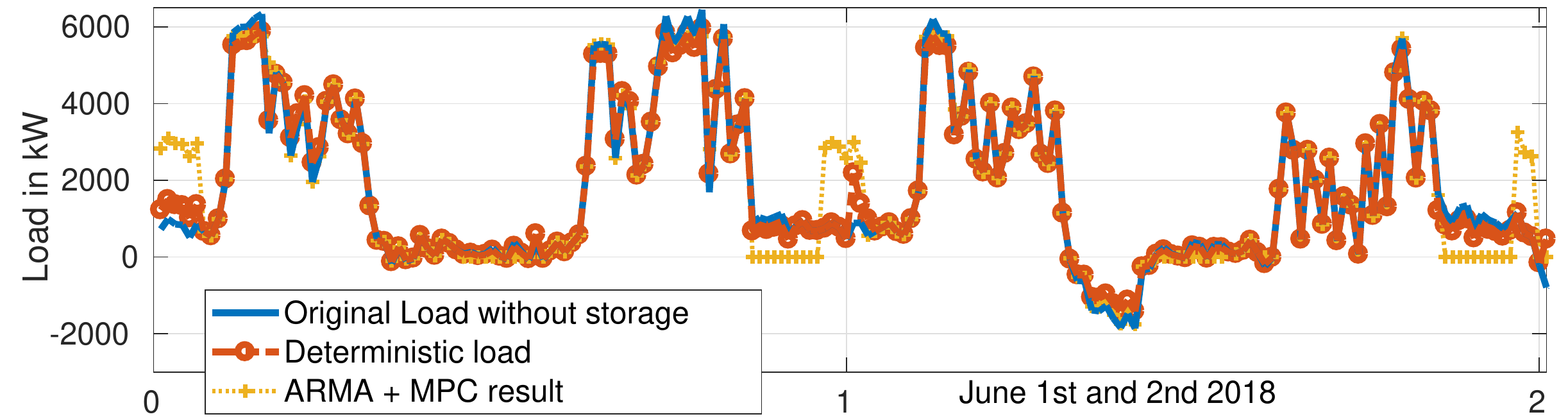}
	\caption{{Net load comparison for June 1 and June 2, 2018}}
	\label{arma2}
\end{figure}

\begin{figure}[!htbp]
	\center
	\includegraphics[width=6in]{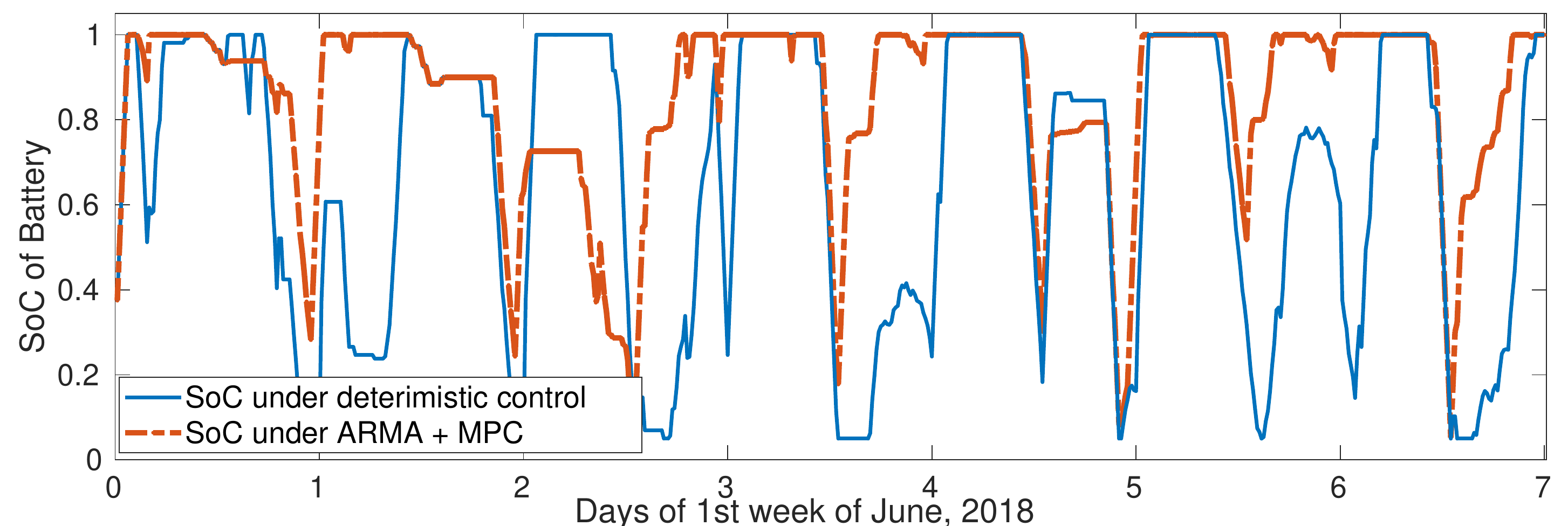}
	\caption{{Battery Capacity for 1st week of June, 2018}}
	\label{armares1}
\end{figure}

The coefficients of the ARMA forecast model is tuned using regression. For this case, we use a battery of 1C-1C type. The comparison of total load for deterministic and MPC simulations is shown in Fig.~\ref{arma2}.
The state of charge (SoC) of the battery is shown in Fig.~\ref{armares1}. It can be observed that for stochastic simulations the battery capacity is maintained at high SoC level in order to minimize probable outage component of the objective function.

The arbitrage gains for the deterministic case for the week is euros 5.50 and for ARMA with MPC the gains are euros 5.01. 
Loss of opportunity (LoO) is defined as = $1-$(actual arbitrage gains)/(deterministic arbitrage gains).
The LoO for this numerical experiment is 8.91\%. A low value of LoO indicates the robustness of our proposed real time framework.
Peak demand shaving for this week is euros 0.336 for both deterministic and ARMA with MPC case. Although in this numerical experiment the peak demand is in compliance with the same contract as for the deterministic case, however, it would be advised to select a higher level of peak demand contract pertaining to forecast errors.

\section{Conclusion}
\label{sectionvi}
We present a case study for the island of Madeira and formulate a convex co-optimization problem for performing arbitrage, peak-shaving and providing backup during power outages. 
Using numerical simulations we observe that storage owners benefit more under greater volatility, evident from higher gains electricity customers can make with 3-level ToU compared to 2-level. 
{We believe an increase in storage size will make more volatile consumer contracts more beneficial.}
Energy storage adds to economic value while solar PV increases self-sufficiency for scenarios where distributed generation is lower or comparable to the magnitude of the inelastic load. For DG generating more than inelastic load, storage also contributes to self-sufficiency by increasing self-consumption.
We show that using storage for power backup during probable and scheduled outages do not undermine its ability to perform arbitrage and peak demand shaving.
{Considering storage operational cycle degradation, we calculated gains per cycle indicating the storage could be financially viable in Madeira (simple payback period $\approx$3 years).}
Numerical simulation for real-time control using ARMA based forecast with MPC shows the efficacy of the proposed scheme for storage co-optimization.
Further work is required to select an optimal battery size based on historical data. 
\section*{Acknowledgments}
{
	Funding from grants: ANR under
	ANR-16-CE05-0008 and EU H2020 under GA 731249 are gratefully acknowledged.}

\bibliographystyle{IEEEtran}
\bibliography{refhard.bib}

\end{document}